# Teaching Economics and Providing Visual "Big Pictures"

Case of General Equilibrium in the IS/LM/AS/AD Framework in Intermediate Macroeconomics[1]

Seyyed Ali Zeytoon Nejad Moosavian
*PhD Student in Economics at North Carolina State University*
Campus Box 8110, NC State University,
Raleigh, N.C. 27695-8110, USA
Phone: +1-206-849-0804
Email: szeytoo@ncsu.edu

---

# Teaching Economics and Providing Visual "Big Pictures"

Case of General Equilibrium in the IS/LM/AS/AD Framework in Intermediate Macroeconomics

Seyyed Ali Zeytoon Nejad Moosavian
*PhD Student in Economics at North Carolina State University*

**Abstract:**

The goal of this paper is to investigate the importance of providing visual "big pictures" in the teaching of economics. The plurality and variety of concepts, variables, diagrams, and models involved in economics can be a source of confusion for many economics students. However, reviewing the existing literature on the importance of providing visual "big pictures" in the process of learning suggests that furnishing students with a visual "big picture" that illustrates the ways through which those numerous, diverse concepts are connected to each other could be an effective solution to clear up the mentioned mental chaos. As a practical example, this paper introduces a "big picture" that can be used as a good resource in intermediate macroeconomics classes. This figure presents twenty-seven commonly-discussed macroeconomic diagrams in the intermediate macroeconomics course, and gives little detail on some of these diagrams, aiming at helping students to get the whole picture at once on a single piece of paper. This macroeconomics big picture mostly focuses on the routes through which common diagrams in macroeconomics are connected to each other, and finally introduces the general macroeconomic equilibrium that is graphically derived through those connections.

**Key Words:** Teaching of Economics, Pedagogy, Undergraduate and Graduate Teaching, Intermediate Macroeconomics, Neoclassical, and Keynesians IS/LM/AS/AD Models.

**JEL Classification:** A22, A23, E12, E13

## 1. Introduction:

The crucial role and growing importance of education in the development of societies are, nowadays, beyond question. Teaching and learning are two essential elements of any educational system. The methods and techniques that instructors employ in teaching have a considerable influence on the quality of learning gained by students. For this reason, pedagogy as the art of teaching and the instructional methods applied by an instructor plays a key role in any educational system, since it is a major determinant of the quality of learning and the level of understanding gained by students.

Economics is a science involving a large number and a wide variety of concepts, variables (which represent the mentioned concepts), and models (which aim to describe the underlying relationships among the mentioned variables). Hence, the proper way of teaching economics could be largely influenced by this distinguishing feature of economics. In order for instructional methods used in teaching economics to be suitable and effective, they need to take this special feature of economics into account so that they can prevent students from possible confusions resulted from this plurality and diversity. The main claim of this paper is that providing a holistic visual "big picture" of the subject matter under discussion can serve as a complementary tactic to take this special feature of economics into consideration. Therefore, a typical big picture can help students remove their probable mental chaos brought about by the plurality and variety of economic concepts, terminology, variables, diagrams, and models through structuring a vivid framework in mind that clearly explains all the underlying relationships among those numerous and various concepts, terminology, variables, diagrams, and models.

Merriam-Webster dictionary defines the term "big picture" as "the entire perspective on a situation or issue, and everything that relates to or is involved in a situation or issue." The term "big picture" points out to the entirety of a situation, avoiding details, all the elements related and involved in a situation, and overall perspectives and views to the issue. The expression "big picture" in this paper is defined as "a visual representation including a set of items (including concepts, variables, diagrams, etc.), which are closely related and logically connected, illustrating how those elements are connected to each other, what types of relationships they have with each other, and in what their origins lie."

Gilbert (2008) states "visual representations exist in two ontological forms. The first of these is as "internal representations" which are the personal mental constructions of an individual, aka mental images. The second of these is as "external representations" which are open to inspection by others" (Gilbert, 2008, p.5). Putting things in accordance with Gilbert's (2008) saying about visual representations, a big picture can also be either in mind (called "mental big picture" or "internal big picture") or on a piece of paper (called "physical big picture" or "external big picture"). In fact, the fundamental idea behind this paper lies in the quote that says "a picture is worth a thousand words", or that "one look is worth a thousand words." More specifically, this quote refers to the notion that a complex idea can be conveyed with just a single image that makes it possible to absorb large amounts of data quickly.

Assessing the existing literature, which is to be carried out in the next section, shows that most of the research studies in the area of economics education have come up with general solutions to the issues with the teaching of economics, such as "general" teaching style modification; however suggesting "specific practical solutions" is still a gap which has not yet been discussed sufficiently in the research area of teaching economics. In fact, this is the gap that this paper tries to fill by proposing a practical solution along with an example practice of that in the real world. Therefore, this paper is not supposed to propose a general teaching style for economics, like what many other authors in the field have done before. Rather, the main aim of this paper is to provide a practical solution to one of the identified problems with the teaching of economics; that is, lack of ability to make connections among so many diverse concepts. In this paper, it is suggested that in many cases the solution can be providing the visual "big picture" of the subject matter under discussion.

The remainder of this paper has been structured as follows. The next section reviews the existing literature related to the topic under discussion. A few of the papers reviewed in this section have been chosen from education literature, and the other reviewed studies are basically classified as the literature of the teaching of economics. Then, the main discussion is offered, through which the reasons why big pictures matter so much in economics, and more specifically in macroeconomics, will be discussed. Afterwards, a practical example of a visual big picture for the general equilibrium in the IS/LM/AS/AD model will be provided. Thereafter, a few considerations which need to be taken into account when designing and applying a big picture are discussed. Finally, a conclusion about the issue will be drawn from the whole discussion.

## 2. Literature Review:

In this section, the existing literature on teaching economics and visualization is reviewed in two separate parts. In the first part, the literature concerning the necessity of paying attention to the ways of teaching economics is reviewed. In the second part, a select group of studies have been reviewed from education literature, primarily concerning the importance of visualization and visual "big pictures" in the process of learning. The information and knowledge gained from this part will be used to build up a theoretical, educational framework to support the main idea of the paper.

### 2.1 Concerning the necessity of attending to the ways of teaching economics

In his article entitled "Goodbye old: hello new in teaching economics", Becker (2004) states that "throughout the world, economists have observed students' lack of interest in pursuing the study of economics" (Becker, 2004, p.1). He then brings up some evidence to support his claim, pointing out to a decrease in percentage US Bachelor's degrees awarded in economics during the past few decades. At the same time, he adds that "there has been an increase in academic economists' interest in their teaching" (Becker, 2004, p.1).

Ongeri (2009) provides a comprehensive survey of the explanations advanced for poor student perceptions of college and university economics teaching and the strategies that have been proposed to reverse these perceptions. He examines the existing literature on the potential causes of poor student evaluations in the teaching of college economics and proposed remedies in order to identify potential strategies for the improvement of teaching in this field as well as directions for future research in this area. According to Ongeri (2009), "research over the last twenty years or so has found that the subject of economics and the quality of instruction in this subject have been consistently ranked among the lowest by undergraduate students in colleges and universities in the United States" (Cashin, 1990; Becker & Watts, 2001b) (quoted by Ongeri, 2009, p.1). In response, he believes, "there has been an increased effort to understand the reasons behind negative student perceptions of economics and what might be done to improve this perception".

In asserting the difficulty that economics students are facing in learning economics, Wilson and Dixon (2009) in an article entitled "Performing Economics: A Critique of Teaching and Learning" state that "economics students find difficulty in developing effective learning strategies. They then verify the difficulty involved in learning economics by saying that "we should be clear, also, that for us the problem is not one of too much abstraction and too little reality. Economics, like other sciences – indeed, like other theoretical discourses – just is abstract, and we and the students had better get used to it. But abstract discourses do not have to be unreal. Economic abstractions have origins, contexts of discovery and actual relations to other principled discourses" (Wilson and Dixon, 2009, p.93). In fact, they are trying to identify some of the missing parts in todays' teaching of economics. They also point out to the fact that the economic abstractions discussed in many of the micro- and macroeconomics textbooks have origins that are not clearly discussed. In addition, they claim that in many of economic textbooks the contexts of discovery and actual relations to other principled discourses are ignored to be discussed. However, they fail to bring up any specific practical solutions to the mentioned problems, and they indeed suffice to introduce the existing problems with teaching economics.

As we have seen in this part of the literature review, a considerable number of research studies have been conducted to address the existing issues involved in teaching economics; however, of those, just a few have come up with "specific practical solutions" to the mentioned problems with teaching economics. In fact, a majority of them have sufficed to either recognize the necessity of change in teaching economics, or introduce and propose some new "general" teaching styles, and they have often failed to suggest some "specific" practical solutions to overcome these issues. More specifically, the lack of practical solutions to overcome difficulties involved in teaching economics is the "gap" or indeed the "missing part" that this paper has found through reviewing the literature, and is to fill the discovered gap somehow within the next few sections.

**2.2 Concerning the crucial roles of visualization and visual "big pictures" in the process of learning**

In emphasizing the importance of representations in the process of learning, Gilbert (2010) states that "representations are the entities with which all thinking is considered to take place. Hence, they are central to the process of learning and consequently to that of teaching. They are therefore important in the conduct and learning of science" (Gilbert, 2010, p.2). Arcavi (2003) examines the role of visual representations in the learning of mathematics. Arcavi (2003) points out that "vision is central to our biological and socio-cultural being" (Arcavi, 2003, p.213). Then, he adds that "the largest part of the cerebrum is involved in vision. The optic nerve contains over one million fibers, compared to 50,000 in the auditory nerve. The study of the visual system has greatly advanced our knowledge of the nervous system. Indeed, we know more about vision than about any other sensory system" (Arcavi, 2003, p.213). Therefore, as biological and as socio-cultural beings, we are encouraged and aspire to 'see' not only what comes 'within sight', but also what we are unable to see." He then refers to a quote from McCormick et al. (1987) stating that "visualization offers a method of seeing the unseen" (Arcavi, 2003, p.216).

In her book called "Teaching at its Best", Nilson (2010) states that structure is so key to how people learn and has such far-reaching implications for teaching. She says without structure there is no knowledge. She believes that "information" is nowadays available everywhere. However, what it is not so available everywhere is organized bodies of "knowledge". She defines knowledge as a structured set of patterns that we have identified through observation. She argues that students are not stupid; they are simply novices in the discipline, who do not see the big picture of the patterns, generalizations, and abstractions that experts recognize so clearly (Arocha & Patel, 1995; DeJoneg & Ferguson-Hessler, 1996). She mentions that without such a big picture, students face another learning hurdle in addition to their other hurdles they may have.

It has been known that the human mind processes, stores, and retrieves knowledge not as a collection of facts, but as a logically organized whole, a coherent conceptual framework, with interconnected parts. Without having a structure of the material in their heads, students fail to comprehend and retain new material (Anderson, 1984; Brandsfor et al., 1999; Rhem, 1995; Svinicki, 2004). The kind of deep, meaningful learning that moves a student from novice toward expert is all about acquiring the discipline's hierarchical organization of patterns, its mental structure of knowledge (Anderson, 1993; Royer, Cisero & Carlo, 1993). "Only then will the student have the structure needed to accumulate additional knowledge" (Nilson, 2010, p.6). She believes instructors should give students the big picture – the overall organization of the course content – very early, and the clearest way to do this is in a graphic syllabus, and instructors should refer back to the visual big picture to show students how and where specific topics fit into that big picture (Nilson, 2010, p.242).

Nilson (2010) points out that "the younger generation of students is not as facile with text as it is with visuals, so a wise idea is to illustrate courses' designs to students so they can see where the course is going in terms of students' learning. Visual aids such as graphic representation of theories, conceptual interrelationships, and knowledge schemata – e.g. concept maps, mind maps, diagrams, flowcharts, comparison-and-contrast matrices, and the like – are powerful learning aids because they provide ready-made, easy-to-process structure for knowledge. Additionally, the very structures of graphics themselves supply retrieval cues (Svinicki, 2004; Vekiri, 2002). As Nilson (2010) reports, according to Kozma et al. (1996), since the chances are very slim that students will independently build such cognitive schemata in a semester of two of casual study, it is wise instructors' task to furnish their students with relevant structure of the associated discipline with valid, ready-made frameworks for fitting the content.

To conclude the literature reviewed here, it should be noted that the existence of a growing amount of literature on teaching economics is a sign of economics instructors' interest in improving their teaching and students' learning in the field of economics. Nevertheless, little attention has been paid to the question of what practical solutions can be employed to resolve these problems. Hence, lack of practical solutions to tackle the introduced issues is still a missing part in solving the issue, and this is indeed what this paper is aimed at doing. In fact, it seems that there is enormous potential with visualization and providing visual "big pictures" of the subject matter being discussed to improve the quality of teaching and learning, economics which has not yet been employed fully in order to solve some of the issues with the teaching of economics.

In the next section, the main discussion of the paper will be offered. The structure of the next section is such that firstly a brief discussion is made about the issue. Then, multiple well-known, widely-used practices in providing big pictures are introduced. Afterwards, a practical example of a typical "big picture" for teaching intermediate macroeconomics is offered, and its specifications and features are explained in brief. Finally, some considerations and tips for designing a typical "big picture" are presented.

## 3. Main Discussion:

### 3.1 The Main Idea

Reviewing the existing literature on the subject of teaching economics indicates that many authors have raised the issues surrounding the teaching of economics. In most cases, authors have sufficed to highlight the issues they have identified (e.g. Becker, 2004). In some other cases, the studies have focused solely on introducing some "general" advice of teaching styles (e.g. Watts, 2001). In fact, the former category deals with "problem identification," and the latter category essentially deals with "structure modification". Although both of these phases of the research attempts are needed and necessary in the process of resolving issues surrounding the teaching of economics, they are not sufficient by any means. These are just the initial steps to take in order to solve such a problem. Indeed, we need to take a step forward, and come up with "specific practical solutions" to resolve the addressed problems that sometimes have their roots in the special features of economics.

Part of the issue has its roots in the plurality and variety of concepts, variables, diagrams, and models involved in economics. This can be a source of confusions for many students majoring in economics. In the event that these numerous and various concepts, variables, diagrams, and models are left scattered in the students' minds without a strong, meaningful framework to connect those apparently separate items, student are most likely to suffer from a potential confusion. However, providing a big picture that illustrates the ways through which those items (concepts, variables, diagrams, and models) are connected to each other could be an effective solution to resolve that problem, and avoid that mental chaos. Providing such a big picture can help students increase their level of comprehension to a large extent.

**3.2 Multiple Typical Examples of "Big Pictures"**

Before going any further, I find it more helpful to first cite multiple well-known, widely-used "big pictures" in economics and its related disciplines such as statistics.

Leemis and McQueston (2008) provide an excellent example of a "big picture" for probability distribution families and their relationships. This "big picture" of distributions not only illustrates the ways through which the distributions are connected, but it also gives some details in notational form to make those relationships clear to audiences. Speaking of microeconomics, Snyder and Nicholson (2012) give a notable example of a big picture linking several demand-related concepts in a single picture. Although when these concepts are introduced one by one and separately they seem to be complicated at first glance to most of the students starting dealing with these ideas, providing them with such an explanatory big picture helps them figure out well how those ideas and concepts are related and connected to each other. The most basic and simplest "big picture" in macroeconomics is in fact the so-called "circulation model". This "big picture" helps students build up a thorough mental framework of an economy as they start studying macroeconomics. Although these big pictures sometimes take different forms, including and excluding various sectors of the economy, they all try to provide the same idea of how a typical economy works. In his book called "macroeconomics", Gartner (2009) has also provided a fine example of a big picture for teaching macroeconomics.

In the next part, a "big picture" for teaching intermediate macroeconomics is introduced.

**3.3 A Practical Example of a "Big Picture" for Teaching Intermediate Macroeconomics**

As mentioned earlier, in many cases, the complaints about the difficulty of learning economics might be a result of missing the big picture. A well-designed visual big picture is capable of helping students to get the whole picture as a mental framework in order to comprehend more easily. A big picture is indeed a zoomed-out version of the course materials of a class that serves not only the students but also the instructors. It deepens students' understanding of the subject as well as that of instructors, enabling them to elaborate further on the material being covered.

Now that we have observed multiple typical examples of big pictures appeared in the areas of economics and statistics, it is time to discuss the important role of a typical "big picture" more specifically in the process of teaching and learning macroeconomics. Considering the existence of a large number of concepts, variables, diagrams, and models in the field of macroeconomics, the significance of having big pictures on paper and consequently in mind is vital in the process of teaching and learning macroeconomics. It becomes even more important when we take into consideration the fact that each of these elements has its own interrelationships with the other elements through one or more channels. Having such a big picture on a paper and subsequently in mind helps students readily deepen their understanding of the topics under study, and also enables them to better analyze these models in the practical situations.

As mentioned before, one purpose of this paper is to discuss the importance of providing and having big pictures in the process of teaching and learning economics as a whole, in general, and to put a great importance on this attempt for the specific case of macroeconomics. A second goal of this paper is to present an obvious example of a holistic "big picture" for general macroeconomic equilibrium in the IS/LM/AS/AD framework in order to provide a practice to show how to design a figure to highlight relationships among related economic concepts and models. In doing so, this paper attempts to demonstrate in what ways a number of various macroeconomic concepts being involved in intermediate macroeconomics are logically interrelated to each other. This task is to be conducted for 27 commonly-arisen diagrams in intermediate macroeconomics. Some, but not all, of the concepts that are supposed to appear in this big picture are as follows: the leisure-work choice problem, labor force supply and demand determination model, aggregate production function and its related features, Solow model, money supply and demand functions, full employment line, user cost of capital model, desired capital stock, Keynesian cross, national saving and investment model, IS/LM curves, aggregate supply and demand, and Philips curve.

Having a chance to look at the whole picture at once and on a single piece of paper is an invaluable opportunity to have for students. At first glance, some of the concepts mentioned above may seem quite irrelevant to each other; however, the truth of the matter is that they are all indeed connected if one looks at the comprehensive, big picture of the subject at hand. The ways through which these key macroeconomic concepts are linked to each other will be shown in the next section. Appendix 1 provides a list of all the abbreviations and symbols used in the following big picture.

# A Visual "Big Picture" for the General Equilibrium in the IS/LM/AS/AD Framework in Intermediate Macroeconomics

## Supply Side

**First Building Blocks / Basic Definitions / Fundamental Ideas**

*The Income/Leisure Trade-off*

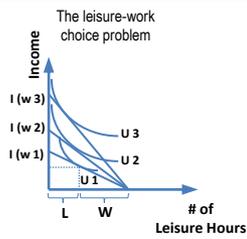

The leisure-work choice problem

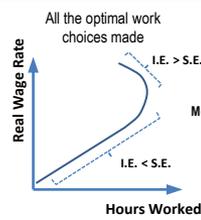

Work-Wage Space → All the optimal work choices made

Most of the time I.E < S.E

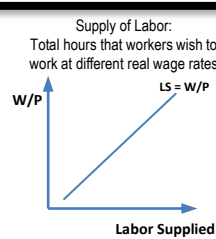

Supply of Labor: Total hours that workers wish to work at different real wage rates

$LS = W/P$

In the Labor Market: Employers hire up to the point at which MPL = MCL

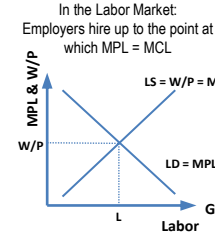

Gives W/P, Gives L

*Diminishing Marginal Product of Labor AND Substitutability of L and K*

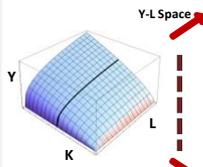

Y-L Space

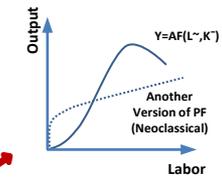

$Y = AF(L^\sim, K^-)$ Another Version of PF (Neoclassical)

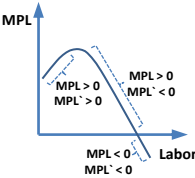

MPL-Input Space, Most of the time MPL > 0, MPL` < 0

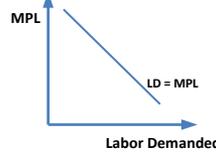

Demand for Labor: Total hours that employers wish to hire workers at different MPL levels

$LD = MPL$

To Define k (Capital Per Worker)

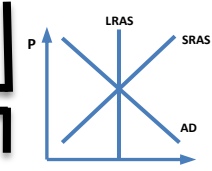

LRAS: $Y = AF(L, K, ...)$  
SRAS: $Y = F(W/P, P/PE, ...)$

Gives LRAS, Gives SRAS

Y-K Space

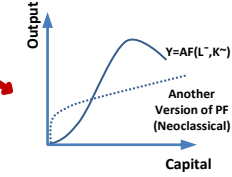

$Y = AF(L^-, K^\sim)$ Another Version of PF (Neoclassical)

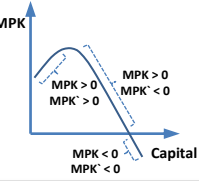

MPK-Input Space, Most of the time MPK > 0, MPK` < 0

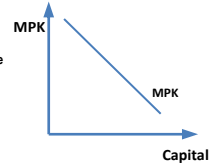

MPK

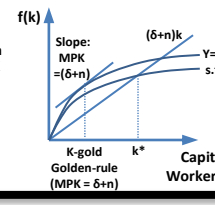

Optimal Capital in Solow Model: The marginal product of capital must equal depreciation plus population growth

Slope: MPK = (δ+n), (δ+n)k, Y=f(k), s.f(k)

K-gold Golden-rule (MPK = δ+n), k*

Idea of per-capita production function & diminishing MPK is used in Solow Model

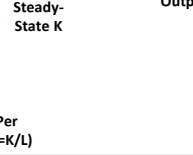

Gives Steady-State K

*Diminishing Marginal Product of Labor AND Substitutability of L and K*

## Demand Side

$MS^- = M0 + M1 + ...$ Assumed to be set by the CB

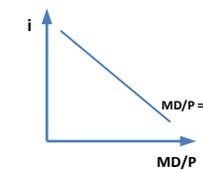

$MD/P = L(Y, i)$

$MD = P \cdot L(Y, i)$ b/c of three functions of money:
1- Medium of exchange (Y)
2- Unit of account (P)
3- Store of value (i)

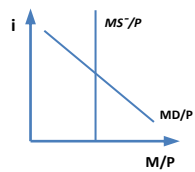

Adding $MS^-$ to the diagram

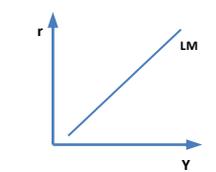

The locus of all equilibria → LM

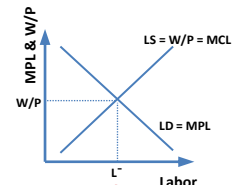

Gives $Y^-$ or Y of FE

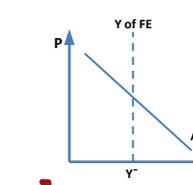

Gives AD

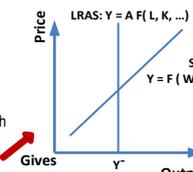

P, LRAS, SRAS, AD, $Y^-$, Y

A Relationship With Phillips Curve

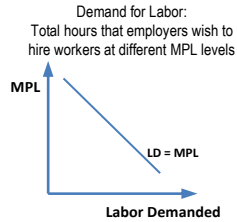

LRPC, SRPC, π, $U^-$, U

*Saving – Real Interest rate relationship is an empirical one.*

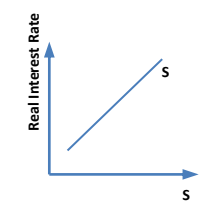

S

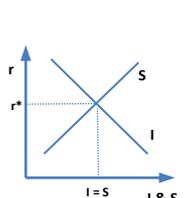

S, I, r*, I = S

The locus of all equilibria. One way to derive IS

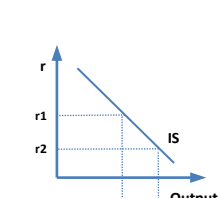

r1, r2, IS, Output

r, Y of FE, LM, IS, $Y^-$, Y

*Interest rate is a part of users' cost of capital*

MPK, UC, UC` = (r`+δ)PK, UC = (r+δ)PK, r↑, K`, K, K & I↓

I-r Space

r, I, Investment

Another way to derive IS: r↓ → I↑

E, E=Y, E = C + I(r2) + G, E = C + I(r1) + G, Y1, Y2, Y, Keynesian Cross

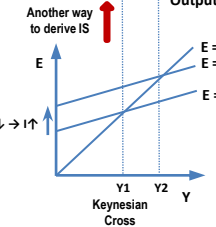

**LM Curve**: The combinations of interest rates and levels of real income for which the money market is in equilibrium  
**OR**: The set of equilibrium points between the liquidity preferences (or MD) function and the MS function

**IS Curve**: The equilibria where total private investment equals total saving  
**OR**: The locus of all equilibria where total spending equals an economy's total output

To see or download a high resolution version of this picture, please go to: http://www.zeytoonnejad.com/macrobigpic.aspx

These relationships are in fact of three types. The first type, which is called "derivative relationship" in this paper, is the case that one diagram is derived from another, or help another diagram be derived in some way. The relationship between IS/LM diagram and AD diagram is of this kind, which constitutes an absolute majority of the relationships in the presented big picture. A second type of relationship among diagrams is the case in which a diagram is the most commonly occurring part of a more complex diagram in the real world. In these cases, when we have to select a part of the curve to keep analysis simple to understand and continue, we usually select the part that is often observed in the real world. For instance, in order to discuss the equilibrium in the labor market we usually pick the upward-sloping part of the labor supply, rather than the downward-sloping part which is believed to be a rare case in the real world. This way of selection is sometimes a source of confusion, and it can be removed by putting a note in the big picture. There is also a third type of relationship among diagrams by which two viewpoints of the same concept are related and connected to each other. An instance of this type of relationship is the relationship between "Keynesian cross" vs. "Classical cross" in the presented big picture. Incidentally, in order to see or download a high resolution version of this picture, you can go to: http://www.zeytoonnejad.com/macrobigpic.aspx

Quite often, Macroeconomics textbooks introduce these models and diagrams in separate sections. Doing so, on the one hand, is advantageous as it helps readers easily concentrate their attention on the subject at hand so that they can better understand what they read. On the other hand, it obscures the understanding of the relationships and linkages among those interrelatedly connected concepts. Failure to provide students with a "big picture" of the concepts being introduced may cause the material to become complicated when the students need to think about the whole picture in order to do an in-depth economic analysis, contributing to weakening the analytical ability of the students. In addition to this, failure to provide students with a big picture of the essentially related economic concepts can also lead to producing students who are lacking critical-thinking features. In other words, failure to provide big pictures to students could cause them to become used to thinking one-sidedly about inherently multi-faceted economic issues while they are looking at solely a part of the picture, missing the other parts of that picture. Such an analysis describes solely a parcel of the whole picture. That is, such analyses look at one aspect of the issue in isolation from the other aspects actually involved.

The big picture presented above is an appropriate visual aid for the teaching of intermediate macroeconomics for eight reasons. First of all, it is a sufficiently comprehensive big picture, including 27 commonly-arisen diagrams in intermediate macroeconomics. Secondly, the "big picture" presented here often points to the types of the relationship among the concepts. Thirdly, the above big picture has been designed in a way that it aims to clear up confusions on different shapes of a single graph. In fact, it shows why sometimes the graph of a single concept takes different forms. Moreover, the big picture presented above sometimes includes some notes necessary to avoid confusion. Further, this big picture includes major definitions, which can be of assistance for student to remember material and avoid confusion. Besides, the above big picture provides fundamental ideas behind some of the diagrams, which are in fact the first building blocks of the diagram being planned to be derived finally. Additionally, this big picture is pluralistic in some parts, providing both versions associated with the Keynesian as well as Classical views simultaneously. Furthermore, the big picture here demonstrates many relationships even beyond the general macroeconomic equilibrium, namely Philips curve. Finally, this big picture labels axes with detailed titles aiming at avoiding confusions caused by lack of information about the precise economic essence of the axes.

Traditionally, economics is taught in a rather routine manner. The role of a typical instructor of economics is usually to introduce topics by giving lectures on general principles, use those principles to derive mathematical models, show illustrative applications of the models, give students practice in similar derivations and applications in homework assignments, and finally test their ability to do the same types of activities and exercises on exams. The approach through which most instructors of economics teach economics is to teach materials chapter by chapter - piece by piece - and when they start a new chapter, they usually prefer, if possible, not to talk about those previous pieces much so that they do not make any confusion for students by relating the new materials under discussion to those of the chapters previously covered. As mentioned briefly earlier, though there is an advantage with doing so, preventing students from being confused by materials of two separate chapters, there is also a major disadvantage, obstructing students from recognizing the interrelationships and linkages among those apparently separate chapters. In order for instructors to have ability to benefit from the advantages of a big picture, and at the same time, to have ability to prevent students from being affected from the respective disadvantage, they can still continue teaching on the basis of "piece by piece" approach, and provide students with "the big picture" at several points in the course repeatedly. By doing so, they help student build a strong mental framework of the underlying connections among the whole materials of the course as they go forward.

A typical big picture should give key points, and emphasize the economic logics among those points. Every concept, example, etc. should be referred to the big picture allowing students visually see what instructors are talking about, and where we are at the moment. Albert Einstein has said: "If you can't explain it simply, you don't understand it well enough." Therefore, every instructor needs to be able to put his or her explanation in simple words and in a big picture, if they themselves understand well what they are teaching. By acquiring an understanding of the big picture and how each idea is logically connected, students become able to unify the concepts. When students can see the whole idea in one picture, they can readily see what the underlying assumptions are, what the causes and effects could be, in what ways the ideas are related to each other, through which channels they are interconnected, and how a single change in the first blocks may have an impact on the last block.

Some authors (e.g. Kennedy, 2008) believe big pictures matters the most at the first stages of teaching. Others may argue that a big picture is a tool that should be used to put things together at the final stages of a course. However, I strongly believe that it can be very helpful at every stage of a course, and it usually becomes more crucial as we move forward toward the final stages of a course. If we take a course as consisting of three time phases, I believe that a big picture can help a class in all the three phases. In the first phase, it can be regarded as "a graphical outline" to illustrate where we are planning to go. Depicting a big picture before diving into in-depth details can help students to grasp the essence of the context. In the middle phase, a "big picture" can be treated as "a road map" or "a broad overview" of the materials being covered in order to demonstrate exactly what and where in the course we are talking about at the moment. Hence, a well-designed "big picture" may be able to prevent students from getting lost during the whole course. Eventually, in the final phase, the big picture can be applied as "a means of putting things together." In other words, big pictures are very useful tools to wrap up a course at the end of the course. This is in fact what helps students achieve the sense of confidence that they have indeed learned something from the course along with a strong mental framework of it. With the passing of the time, many parts of the material learned in a course might be gone; however, the thing that can remain in mind forever or at least on a piece of paper to be reviewed fast if necessary, enabling students to remember the principles of the subject quickly, is a visual "big picture."

Providing a big picture can have so many advantages not only for students but also for instructors. Many of these advantages can serve as a means of solving some of the addressed issues with the teaching of economics. Some, but not all, of the advantages discussed in this paper are as the following: creating a strong mental framework of the subject, preventing students from getting lost among various materials, increasing comprehension level, enabling students to conduct holistic analyses, allowing instructors to take a multimodal approach to teaching, providing a tool to discuss pluralistic ideas in teaching economics, etc.

Needless to say, a typical big picture ignores a large amount of details; however, this is indeed its philosophy to be so. That is, the mission of a big picture is to retain the major ideas, and demonstrate the ways through which those major concepts are connected to each other. Therefore, a big picture serves as the framework of a course, and the lecture notes, lectures themselves, textbook, and other sorts of the materials instructors typically take advantage of in classes will provide the required details to deepen the students' understanding of the material being covered.

The next section of the paper draws a conclusion of the whole discussion offered in this paper.

## 4. Conclusion:

Economics is a science involving a large number and a wide variety of concepts, variables, diagrams, and models. This plurality and variety can be a source of confusion for many economics learners. As a consequence, the proper way of teaching economics could be largely influenced by this distinctive feature of economics. In order for instructional methods employed in teaching economics to be suitable and effective for this specific science, they must take this special feature of economics into account so that they prevent students from possible confusions resulted from this plurality and diversity. This paper suggests that providing a holistic visual "big picture" of the subject matter under discussion can serve as a complementary tactic to take this special feature of economics into consideration in the process of teaching. Therefore, a typical big picture can help students avoid probable mental chaos caused by the plurality and variety of economic concepts, terminology, variables, diagrams, and models through structuring a vivid framework on a paper and subsequently in mind. Such a "big picture" clearly explains all the underlying relationships among those numerous and various concepts, terminology, variables, diagrams, and models. What big pictures potentially can do is to help students clear up the fuzziness and confusions about the scattered materials being covered.

Assessing the existing literature on the teaching of economics shows that most of the research studies in this area have come up with general solutions to the identified issues, but suggesting "specific practical solutions" is still a gap which has not yet been discussed sufficiently in the research area of teaching economics. In fact, this is the gap that this paper tries to fill by proposing a practical solution along with a practical example of that in the real world. As a practical example, this paper introduces a "big picture" that can be used as a resource in intermediate macroeconomics classes. This figure presents 27 commonly-arisen macroeconomic diagrams in the intermediate macroeconomics course, and gives little details on some of those diagrams, aiming at helping students to get the whole picture at once on a single piece of paper, and mostly focuses on the ways through which those diagrams are connected to each other, and finally introduces the general macroeconomic equilibrium. This "big picture" ends up with general macroeconomic equilibrium in the IS/LM/AS/AD framework. In doing so, this paper attempts to demonstrate in what ways a number of various macroeconomic concepts being involved in intermediate macroeconomics are logically interrelated to each other.

The final point of this paper is the fact that no matter what courses of economics an economics instructor is teaching, economics instructors should not leave the structure they are building in their students' minds without a strong framework, which will be indeed their visual "big picture." Scattered pieces of information in minds do not stay there for long; and even if they did, they would not be useful to stay there. Economics instructors can design their own visual big pictures according to their teaching experiences, preferences, thinking, etc. They can also bring it up in different phases of their class, whenever they prefer to. After all, what they should not or cannot do is to leave their students without a "big picture" finally, and if they do so, they will be leaving the students with the nightmare of lacking a mental framework of the material covered.

# Appendix 1: Symbols and Notations

**Symbols:**
¯ : Fixed
~ : Changing

**Notations**:

I: Income
I(w1): Income at wage rate 1
U: Utility level OR Utility Indifference Curve
L: Leisure hours
W: Hours worked
W: Nominal wage rate
w: Real wage rate OR W/P
P: Price level
I.E: Income effect
S.E: Substitution effect
L: Labor supplied OR Labor hours worked
LS: Supply of labor
LD: Demand for labor
MCL: Marginal cost of labor
MPL: Marginal product of labor
Y: Output (Income)
A: Technology level OR Total Factor Productivity (TFP) level
PF: Production function
MPL': Derivative of marginal product of labor with respect to L
MPK: Marginal product of capital
MPK': Derivative of marginal product of capital with respect to K
F(.) OR f(.): Function of
$\delta$: Depreciation rate
k: Capital per worker (K/L)
k*: Steady-state k
k-gold: Golden-rule k
n: Population growth rate
s: Saving rate (S/Y)
LRAS: Long-run aggregate supply
SRAS: Short-run aggregate supply
AS: Aggregate supply
AD: Aggregate demand
FE: Full employment
Y¯ of Y of FE: Output level at full employment
PE: Price expectation
MS¯: Money supply
M0: Sum of currency in circulation (notes and coins) plus banks' reserves with the central bank
M1: Currency in circulation plus current (checking) accounts plus deposit accounts transferable by checks
i: Nominal interest rate
r: Real interest rate
MD: Money demand (Demand for money)
LM: Liquidity-Money equilibrium curve
L(Y, i): Liquidity function
S: National saving
I: National investment
IS: Investment-Saving curve
K: Capital stock

UC: User cost of capital
E: Expenditures
G: Government Expenditures
I(r1): Investments made at the interest rate "r1"
C: Consumption
$\pi$: Inflation rate
U: Unemployment rate
U¯: The natural rate of unemployment
LRPC: Long-run Philips curve
SRPC: Short-run Philips curve

**Appendix 2: List of the Diagrams and Models in the Visual Big Picture of Macroeconomics**

1. The Leisure-Work Choice Problem
2. Individual Labor Supply Curve
3. Labor Supply Diagram
4. Two-dimensional Production Function Diagram (Y-L Space)
5. Marginal Product of Labor (MPL) Diagram
6. Labor Demand Diagram
7. Labor Market Equilibrium Diagram
8. Three-dimensional Production Function Diagram (Y-L-K Space)
9. Two-dimensional Production Function Diagram (Y-K Space)
10. Marginal Product of Capital (MPK) Diagram
11. Capital Demand Diagram
12. Solow Model
13. Aggregate Supply (AS) Diagram
14. A Diagram for General Equilibrium in the Macroeconomy
15. Money Demand Diagram
16. Money Market Equilibrium Diagram (Money Supply and Demand)
17. LM Diagram (Liquidity-Money Diagram)
18. Labor Market Equilibrium Diagram
19. Aggregate Demand (AD) Diagram
20. Phillips Curve
21. Saving vs. Interest Rate Diagram
22. National Saving and Investment Model (aka "Classical Cross" Model)
23. IS Diagram (Investment=Saving Diagram)
24. IS-LM Model
25. User Cost of Capital Model
26. Investment vs. Interest Rate Diagram
27. Aggregate Expenditure Line (aka "Keynesian Cross" Model)